\documentstyle[aps,twocolumn]{revtex}
 \begin{document}

\draft

 \title{Why stripes? Spontaneous formation of inhomogeneous structures
 due to elastic interactions}

 \author{D.I. Khomskii $^{1}$ and K.I. Kugel $^{2}$}
 \address{$^{1}$ Laboratory of Solid State Physics,
 Groningen University, Nijenborgh 4,
 9747 AG Groningen, The Netherlands \\ $^{2}$ Institute for
 Theoretical and Applied Electrodynamics, Russian Academy of Sciences,
 Izhorskaya Str.\ 13/19, 127412 Moscow, Russia}

 \maketitle

 \begin{abstract}

 We argue that elastic interactions between ions in different valence
 states can play an essential role in stabilization of stripes
 (or $2D$ ``sheets'') in doped oxides. These interactions are in general
 long-range and an\-iso\-tro\-pic
 (attractive in certain directions and repulsive in others). This
 can naturally give rise to stripe-like structures in insulating 
 materials.
 We illustrate this general idea with certain specific examples
 and show that the situation can be described by the 
 Ising model  with anisotropic interactions. The case of anisotropic
 impurities, relevant e.g. for manganites, is also briefly discussed.
 \footnote{Accepted for publication in Europhysics Letters}

\end{abstract}

 \tighten
 \bigskip

 The notion of stripes became recently one of the most important 
 concepts in the physics of doped Mott insulators. There are many
 theoretical ~\cite{zaanen&gunnarson,scalapino,stoik} and
 experimental ~\cite{cheong,radael,chen,tranquada} indications
 that they indeed arise in doped transition metal oxides;
 there exist also serious arguments that they may play an important
 role in high-temperature superconductivity
 ~\cite{zaanen&gunnarson,emery,seibold}. 

Despite very intense studies in this field, the origin of
 stripes in real materials is far from being well understood. Following
 the pioneering work ~\cite{zaanen&gunnarson} the main attention
 is paid to the purely electronic and magnetic mechanisms of stripe
 formation ~\cite{scalapino,stoik}. However the existence of 
 stripes in this approach is still controversial ~\cite{hellberg}.

When discussing the origin
 of stripes in doped oxides, one usually starts with Mott insulators
 with an integer number of electrons per site forming local magnetic
 moments. Generically their ground state is antiferromagnetic.
 In this case, one argues that with doping a superstructure may develop
 consisting of antiferromagnetic domains, the doped charge carriers being
 localised at the domain walls. This reflects the general tendency:
 instability of a homogeneous state in doped strongly correlated
 electron systems toward phase separation ~\cite{vissch,nagaev,moreo,khom}.
 Stripe phases are possible manifestations of this tendency.

 Why are stripe phases better than any other forms of phase
 separation? This is the main question, which, in our opinion, still
 did not find a satisfactory answer. One can argue that in systems
 with only short-range interaction, such as the conventional Hubbard
 or $t$--$J$ models, one would rather expect a total phase separation
 into two phases: an antiferromagnetic insulator without any holes ($n=1$)
 and another phase---metallic and probably ferromagnetic---containing
 all the holes; this state allows to gain the full kinetic energy
 of holes, and simultaneously it costs a minimum of surface energy.

 To stabilize the stripe phase, one often invokes long-range
 Coulomb interaction which prevents the large-scale phase separation
 ~\cite{emery,seibold}.
 However, for the strong long-range repulsion we would rather expect
 not stripes but a sort of Wigner crystal, in which the doped carriers
 keep as far away from each other as possible. In this case, stripes,
 and even more so two-dimensional sheets observed e.g. in 
 manganites~\cite{cheong,radael}, are definitely less
 favourable. One then argues that stripes would be stable in the
 intermediate case. And although this scenario is of course not
 excluded, it seems that some important physical factors are
 missing from such a picture. \footnote{Note also that in nickelates
 and manganites, in contrast to cuprates, the metal-centered stripes
 would not correspond to the antiferromagnetic domain walls inherent
 to the conventional theories
 \cite{zaanen&gunnarson,scalapino,stoik,hellberg}.}

 We argue that there is indeed one such factor, which exists in all
 real systems and which can, in principle, stabilize the stripe and
 sheet phases. This factor is the long-range elastic interaction.
 We consider below the insulating materials (nickelates,
 overdoped manganites) with the so called filled stripes
 (one hole per site); the situation in cuprates with presumably
 half-filled stripes may be more complicated, although we believe
 that the factors invoked below can also play an important role there.

 When we dope a
 stoichiometric system, e.g.\ substitute Ca for La in LaMnO$_3$, we
 create Mn$^{4+}$ ions in the Mn$^{3+}$ matrix. Or, in the overdoped
 La$_{1-x}$Ca$_x$MnO$_3$ with $x > 0.5$ (this is the
 situation, in which the stripe, or rather sheet phases were
 observed in manganites~\cite{cheong,radael}), one can speak of a certain
 amount of Mn$^{3+}$ ions in the Mn$^{4+}$ matrix.

 Classically, Mn$^{3+}$ ions differ from Mn$^{4+}$
 (or Ni$^{2+}$ from Ni$^{3+}$ in nickelates~\cite{chen,tranquada})
 not only by their charge, but also by their ionic radius.
 This factor is usually ignored in the conventional treatment of
 stripes. In contrast, we pay the main attention here just
 to this factor (which, of course, should be considered alongside
 with the other factors, usually accounted
 for~\cite{zaanen&gunnarson,scalapino,stoik,hellberg,emery,seibold}).

 As is known from the theory of elasticity, when we create a lattice
 distortion (in simple case---a dilatation, e.g.\ by cutting out a
 small sphere of radius $A$ and substituting it by the sphere of radius
 $A'\neq A$), we also create a field of lattice strains, which is
 long-range and, in general, anisotropic~\cite{eshelby,khach}.
 This leads to a long-range interaction between impurities ($\sim1/R^3$),
 which depends on the elastic constants of the medium and on the shape
 of impurities.

 The most important point is that typically this interaction
 resembles a quadrupole--quadrupole interaction, being
 repulsive in certain directions but attractive in others. This
 long-range attraction opens a possibility to form
 regular structures---clusters of impurities forming one-dimensional
 objects (needles or stripes) or two-dimensional structures (lamellae
 or sheets); this can be one of the mechanisms stabilizing stripe or
 sheet phases.

 The interaction of dilatation impurities in crystals 
 vanishes in isotropic media, but it is nonzero in crystals. 
 Thus, in crystals with cubic anisotropy
 it is given by~\cite{eshelby}
 \begin{equation}
 V(\vec r, \vec r')=-Cd\, Q_1 Q_2 \frac{\Gamma(\vec n)}{|\vec r-\vec
 r'|^3}
 \label{eq1}
 \end{equation}
 where $C$ is a constant of the order of unity, $\vec r-\vec r'=|\vec r-\vec
 r'|\cdot\vec n$, \ $Q_1$ and $Q_2$
 are the ``strengths'' of impurities ($Q_i\sim(v_i-v_0)$ where $v_i$
 is the volume of the impurity and $v_0$ is the corresponding volume
 of the matrix), and
 \begin{equation}
 d=c_{11}-c_{12}-2c_{44},
 \label{eq2}
\end{equation}
 where $c_{ij}$ are the elastic moduli of the crystal. The angular
 dependence of the interaction~(\ref{eq1}) is determined by a
 function of the direction cosines of the vector
 $\vec R=\vec r_1-\vec r_2$:
 \begin{equation}
 \Gamma(\vec n)=n_x^4+n_y^4+n_z^4-\frac35.
 \label{eq3}
 \end{equation}
 From (\ref{eq1})--(\ref{eq3}) we can see that the interaction between
 impurities is long-range ($\sim 1/R^3$) and has different
 signs in different directions. Thus, for the situation with
 $d>0$ it is attractive along [100], [010] and [001] directions
 ($\Gamma([100])=\frac25$) and is repulsive along [110], [011] etc.\
 ($\Gamma([110])=-\frac{1}{10}$) and along the cube diagonals~[111]:
 $\Gamma([111])=-\frac{4}{15}$, and vice versa for $d<0$.
 Thus, if we put a few dilatation impurities in a cubic crystal,
 they will attract each other along certain directions 
(e.g.\ [100], [010], [010]) and repel along others,
 thus causing a formation of inhomogeneous structures (stripes,
 sheets etc.)\footnote{Note also that in contrast to the Coulomb forces,
 these elastic interactions are not screened (except by
 the ``mirror forces'' due to the surface of the crystal~\cite{eshelby}).}

 In real systems (not the weakly anisotropic crystals considered
 above as an example), the ratios of the interaction constants between
 different neighbors may differ from those given above; thus, e.g.,
 in bcc iron the interaction with [011] neighbor is stronger than
 with [001] one~\cite{khach}. But again, as in the example treated
 above, the interactions have different signs in different directions.
 Thus, we can consider the situation of a cubic lattice with the 
 nearest-neighbor ($nn$) and next-nearest-neighbor ($nnn$) interaction
 of opposite signs as a typical one, treating
 the ratio of these couplings as an arbitrary parameter.
 This naturally leads to the lattice gas, or the Ising model of the type
 \begin{equation}
 H=-J\!\!\sum_{\langle ij\rangle=nn}\!\!\sigma_i\sigma_j
  + J'\!\!\sum_{\langle ij\rangle=nnn}\!\!\sigma_i\sigma_j+\rm const
 \label{eq4}
 \end{equation}
 where $\sigma_i=\pm1$; the density at site $i$ is
 $n_i=\frac12(1+\sigma_i)$, and the model has to be considered for 
 fixed density $n=\langle n_i\rangle$ (or fixed ``magnetisation'').
 We treat below the case of a cubic lattice.

 For $J$, $J'>0$ (which corresponds to
 $d>0$ in Eq.~(\ref{eq1})) we have the following solutions:

 1. Totally phase separated state, in which all the ``impurities''
 (or electrons) form one big cluster. The energy of this state
 (per impurity) in a simple cubic lattice is
$E_{\rm ph.sep.}=-Jz_{nn}/2+J'z_{nnn}/2 =-3J+6J'$.
Here $z$ are the corresponding numbers of nearest and next nearest
 neighbours.

 2. Occupied sites forming $2D$ sheets parallel to $(xy)$, $(xz)$ or
 $(yz)$ planes, such that the distance between them exceeds one lattice
 spacing. The energy of such state is 
$E_{\rm sheet}=-2J+2J'$.

 3. The $1D$ stripes in $x$, $y$ or $z$ directions, which
 do not cross or approach one another to one lattice
 spacing. The energy of such a state is
$E_{\rm stripe}=-J$.

 One can easily show that all the other possible structures have higher
 energy.

 By comparing the energies of these states, we see that the totally
 phase separated state will be stable if 
 $J'/J<\frac14$; for $\frac14<J'/J<\frac12$ the 2D ``sheets'' are stable,
 and for $J'/J>\frac12$ the 1D stripes would give a minimum of the
 energy. Thus, we see that the sheet or stripe phases 
 appear quite naturally if the dominant interaction is the
 elastic interaction  between impurities which is intrinsically
 anisotropic and long-range. For $d<0$, we similarly obtain, instead 
 of the vertical, the diagonal stripes, but in this case, we should also 
 include the interaction with third neighbors.

 For low electron (or impurity) density in 3D case the lamellae phase
 is formed by parallel sheets; the distance between them in model
~(\ref{eq4}) with only $nn$ and $nnn$ interactions is arbitrary,
 provided only that it exceeds the lattice spacing. Longer-range
 interactions  will make this arrangement regular, but for weak
 interaction of this kind the discommensurations could be easily
 formed.

 The situation is a bit more complicated for the stripe phase. In
 layered materials, it will be similar to that with the sheets in $3D$
 crystals (parallel equally spaced stripes). In $3D$ crystals
 there may in principle exist mutually perpendicular stripes.
 Again, one may expect that longer-range interaction can
 make these stripes parallel; this question, however, goes beyond the
 scope of our treatment. Thus, we see that the elastic interaction of
 the ``spherical'' impurities in crystals may quite naturally lead to
 the formation of inhomogeneous structures---$1D$ stripes, $2D$ sheets
 or $3D$ phase-separated clusters, which can be described by 
 the Ising-like model~(\ref{eq4}) \cite{kusm}. Note also that this process
 does not require the atomic diffusion: it is realized by the electron
 hopping between, say, $Mn^{3+}$ and $Mn^{4+}$ (with the corresponding
 lattice relaxation).

 The problem of formation of regular structures, when we have two
 components with different atomic volumes, is well-known in the
 physics of segregating alloys. Theoretical
 studies (see e.g.~\cite{khach}) have shown that the shape of inclusions
 of one phase in another is to a large extent determined by the
 elastic strains: the shape is such as to minimize the total
 elastic energy. As is shown in the corresponding studies, the best
 way is to create the second phase in the form of infinitely thin
 layers having certain specific orientation in given matrix. In
 the corresponding alloys it gives rise to the so called
 Guinier--Preston zones~\cite{preston,guinier}: inclusions of the second
 phase appear as a regular array of parallel platelets. One can
 speculate that the physics of the stabilization of stripe phases
 (actually sheets) in La$_{1-x}$Ca$_x$MnO$_3$ at $x = 0.67$, 0.75,
 etc.~\cite{cheong,radael} is the same, and that it is just
 these long-range anisotropic elastic forces
 which stabilize such structures. This is supported by the fact that
 this stripe charge ordering is obtained in the insulating materials
 at temperatures above the eventual magnetic ordering; thus, the often
 invoked magnetic mechanisms play here a minor role, if any.
 The inclusion of the electron hopping will, of course, tend to
 destabilize these ordered structures, which
 could survive if the hopping integrals are not too large.

 The situation may be different (and actually much richer) if the
  ``impurities'' are not simple dilatation centers (``spheres''), but
 are anisotropic (``ellipsoids'').  This is the typical situation with
 Jahn-Teller ions, e.g.\ Mn$^{3+}$ ``impurities'' in overdoped
 manganites R$_{1-x}$Ca$_x$MnO$_3$ (R = La,~Pr,~Nd) for $x>0.5$. 
 The interaction between such impurities also decays as $1/R^3$ and
 has different signs depending on both the relative position of such
 centres in a crystal and on the orientation of corresponding orbitals
 (i.e.\ local distortions)~\cite{erem,fishman}. Thus, for example,
 for two quadrupolar ions along the $z$-axis with the electron
 density elongated parallel to $x$, $y$, or $z$ axes, as is the
 case of $e_g$-ions like Mn$^{3+}$
 one obtains from the general expressions \cite{erem} 
 the interaction in the form
 \begin{eqnarray}
 V&=&\frac{(c_{11}+c_{44})}{8\pi(c_{11}+2c_{44})R^3}\Bigl\{5\sigma_{zz}^{(1)}
 \sigma_{zz}^{(2)}+ \nonumber\\
 &&+2\Bigl(\sigma_{xx}^{(1)}\sigma_{xx}^{(2)}+\sigma_{yy}^{(1)}
 \sigma_{yy}^{(2)}\Bigr)\Bigr\}+\nonumber\\
   &&+\frac{1}{4\pi R^3}
 \Bigl(2\sigma_{zz}^{(1)}\sigma_{zz}^{(2)}-\sigma_{xx}^{(1)}\sigma_{xx}^{(2 )}
 -\sigma_{yy}^{(1)}\sigma_{yy}^{(2)}\Bigr)
 \label{eq8}
 \end{eqnarray}
 where $c_{11}$ and $c_{44}$ are elastic moduli, and
 $\sigma_{\alpha\alpha}$ is a stress tensor such that e.g.\ for the
 center with the occupied orbital $3z^2-r^2$ we have $\sigma_{zz}=1$,
 $\sigma_{xx}=\sigma_{yy}=-1/2$ (and corresponding expressions for
 $z \to x, y$).

 From this expression one can see that e.g.\ two $3z^2-r^2$ orbitals along
 $z$-axis strongly repel one another, whereas $3z^2-r^2$ and $3x^2-r^2$
 (or $3y^2-r^2$) along the same direction attract. This can quite
 naturally explain the well-known orbital structure of undoped
 LaMnO$_3$ (cf.\ \cite{novak}), and also a special stability of the
 Mn$^{3+}$-planes, with the corresponding orbital ordering, in lightly
 doped manganites, e.g.\ La$_{1-x}$Sr$_x$MnO$_3$, $x\sim1/8$
 \cite{mizokawa} and in Pr$_{1-x}$Ca$_x$MnO$_3$,
 $x\sim1/4$~\cite{hill}. With the assumption of
 a checkerboard charge ordering it also gives the orbital
 ordering of CE type observed in R$_{1-x}$Ca$_x$MnO$_3$ 
 at $x=0.5$~\cite{wollan,jirak}. Indeed, in both cases the
 orbitals of the neighbors along $x$ and $y$ directions are mutually
 orthogonal (e.g.\ $3x^2-r^2$ and $3y^2-r^2$ orbitals); according to
 ~(\ref{eq8}) they have an attractive interaction which stabilizes
 these structures.
 One can also show that the interaction along diagonals ([110] direction)
 is attractive for the parallel orbitals $3x^2-r^2$ or $3y^2-r^2$ but
 repulsive for the perpendicular ones
 ($3x^2-r^2$ at one site and $3y^2-r^2$ at the other);
this gives extra stability to the observed orbital ordering in undoped
LaMnO$_3$, and these diagonal interactions nearly cancel for the
 CE-type phase in the $x=0.5$ system. The same factors may favor
 single stripes ("Wigner crystal")~\cite{radael} as compared to 
 the bi-stripes~\cite{cheong}.

 Summarizing, we suggest that the elastic interactions, always present
 in systems with mixed valence, e.g.\ in doped Mott insulators, may
 play an important role in stabilizing particular types of
 inhomogeneous structures, such as stripes or two-dimensional sheets.
 These interactions decay rather slowly ($\sim R^{-3}$) and
 are anisotropic, being repulsive in some directions but attractive in
 others. Thus, the interaction of dilatation impurities in
 weakly anisotropic cubic crystals has different signs along [100],
 [010], and [001] directions and along face or body diagonals. Such an
 interaction may quite naturally lead to the formation of
 stripes or sheets in manganites or nickelates. Similar
 interactions in the case of anisotropic (Jahn-Teller) ions can
 produce analogous inhomogeneous structures with
 specific orbital ordering. The importance of lattice effect in
 the stripe formation is supported by the observation of a large
 isotope effect on the stripe ordering temperature~\cite{lanzara,rubio}.
 Thus, we can conclude that the elastic or, more generally,
 electron--lattice interactions may be very important in
 providing a mechanism for stripe formation in doped systems.

We are grateful to S.-W.~Cheong, M.~V.~Eremin, A.~Ya.~Fishman,
 J.~Hill, S.~Kivelson, F.~Kusmartsev, V.~Ya.~Mitrofanov, and
 G.~Sawatzky for very useful discussions.

 The work was supported by INTAS (grants 97-0963 and 97-11954),
 the Russian Foundation for Basic Research (grant 00-15-96570),
 the Netherlands Foundation for the Fundamental Research of
 Matter (FOM), and the Netherlands Organization for the
 Advancement of Pure Research (NWO).

 \end{document}